    \documentclass{elsart}

     \usepackage{graphicx}
      \include{epsfig}
    \usepackage{amssymb}
        \usepackage{amsmath}
    
\begin{document}
    \begin{frontmatter}


    \title{Statistical auditing and randomness test of lotto $k/N$-type games}
    \author{H.F. Coronel-Brizio},
    \ead{hcoronel@uv.mx}
    \author{A.R. Hern\'andez-Montoya},
       \ead{alhernandez@uv.mx}
    \address{Facultad de F\'{\i}sica e Inteligencia Artificial.
    Universidad Veracruzana, Sebasti\'an Camacho 5, Xalapa, Veracruz 91000. M\'{e}xico}
    \author{F. Rapallo},
    \ead{fabio.rapallo@mfn.unipmn.it}
    \author{E. Scalas}
    \ead{enrico.scalas@mfn.unipmn.it}
    \ead[url]{www.mfn.unipmn.it/\textasciitilde scalas}
    \address{Dipartimento di Scienze e Tecnologie Avanzate, Universit\`a del Piemonte Orientale, Via Bellini 25/G, 15100 Alessandria, Italy}

\begin{abstract}

One of the most popular lottery games worldwide is the so-called
``lotto $k/N$''. It considers $N$ numbers $1,2,\ldots,N$ from
which $k$ are drawn randomly, without replacement. A player
selects $k$ or more numbers and the first prize is shared amongst
those players whose selected numbers match all of the $k$ randomly
drawn. Exact rules may vary in different countries.

In this paper, mean values and covariances for the random
variables representing the numbers drawn from this kind of game
are presented, with the aim of using them to audit statistically
the consistency of a given sample of historical results with
theoretical values coming from a  hypergeometric statistical
model. The method can be adapted to test pseudorandom number
generators.
\end{abstract}
\begin{keyword}
Empirical Randomness Test\sep Lottery Games \sep Multivariate Hypothesis Testing \sep Combinatorial Calculus \sep Statistical models
\PACS 01.75.+m \sep 02.50.Cw \sep 02.50.Ng \sep 02.50.Sk \sep 07.05.Kf \sep 89.20.-a \sep 89.65.Gh \sep
\end{keyword}
\end{frontmatter}

\section{Introduction}
\label{introduccion}

The concept of chance has a long history, but, as far as we know,
early scientists and mathematicians working in the Hellenistic
period did not develop either a theory of probability or
statistical methods \cite{hacking}. Gambling became more and more
popular in Europe in the XVIIth century, due to the emergence of a
class of people affluent enough to travel along the continent and
waste money in such games. Games of chance were at the origin of a
new wave of interest on the {\it rules} of chance \cite{hacking}
and fostered the first rigorous results in Probability Theory.
Among all the games of chance, lotteries have been and still are
very popular. They are used by governments to levy indirect taxes
on very poor people. It is not clear when the first European
lottery games started, but it seems that they could have been
already present in the XVth century. Influential names in the
history of science, such as D'Alembert, Euler, D. Bernoulli,
Huygens, Leibniz, Laplace and many others analyzed lotteries for
practical purposes, such as designing them and optimizing
governmental collected revenues, but also with theoretical goals
in mind, helping to accelerate the development of Statistics and
Probability Theory. A very interesting account on the history of
lotteries emphasizes the role of Genoa (an Italian Sea Republic of
the Middle Ages) in introducing state-run lotteries \cite{Genoa}.
That paper includes further interesting references.

Nowadays, analysis, design  and simulation of lottery games
continue to be an active research area, mainly  for statisticians
and economists \cite{today1,today2,today3,today5}, and also
studied as a suitable tool for teaching elementary probability
theory and Statistics \cite{education1,education2}, but even new
interesting theoretical results have been obtained recently
\cite{stephens}.

In this work we present a statistical data analysis of randomness
of Mexican and Italian lotteries; although, strictly speaking, it
is known that there is no way to ``prove'' the randomness of a
sequence of numbers \cite{chaitin}, it is always possible to
statistically test whether or not historical results exhibit the
quantitative properties derived from the probabilistic model
assumed to explain the selection mechanism. In this respect, the
statistical procedure presented here  could be easily used as a
test of pseudorandom number generators.

\section{Theory}

\subsection{Probabilities}

Readers can find the following references useful to understand the
material presented in this section: \cite{Feller,Prob1,Comb1} for
what concerns Probability and combinatorial analysis and
\cite{lehman} for Statistics.

The total number of possible combinations of $k$ objects chosen
from a set of $N$ objects is given by the combinatorial
coefficient ``$N$ choose $k$'':
\[
\dbinom{N}{k} = \frac{N!}{k!(N-k)!}.
\]
We denote by $X$ the random variable corresponding to the number
of matches out of the $k$ randomly drawn numbers. Here, we use the
hypergeometric model and we prefer the technical term ``fairness''
in place of ``equiprobability'' as, strictly speaking, all the
lottery numbers are equivalent-exchangeable, but the odds of
extracting them do not follow the uniform distribution (sampling
is without replacement) and, in drawing each of the $N$ objects,
the probability of matching exactly $i$ numbers, out of $k$
selected by the player, is given by \cite{definetti}
\begin{equation} \label{prob}
P\left[ {X = i}\right] =
\dbinom{k}{i}\dbinom{N-k}{k-i}\dbinom{N}{k}^{-1}
\end{equation}
where $i=0,\ldots,k$.

In order to test the hypothesis of fairness, we consider a
multivariate test on the mean parameter of the random variable
${\bf Y'} = \left[ {Y_{\left( 1 \right)} , \ldots ,Y_{\left( k
\right)} } \right]$, the {\em sorted outcome vector}. Here,
$Y_{\left( i \right)}$ denotes the random variable corresponding
to the number in the $i-$th place (recall that the randomly
selected numbers are put in ascending order i.e., $Y_{\left( 1
\right)}  < Y_{\left( 2 \right)}  < \ldots  < Y_{\left( k
\right)}$). The probability that the $i-$th number corresponds to
the value $r$, is calculated from Eq. \eqref{prob} with a suitable
choice of parameters. In fact, $Y_{\left( i \right)} = r$ if and
only if $i-1$ numbers fall between $1$ and $r-1$, and $k-i$
numbers fall between $r+1$ and $N$. Therefore,
\begin{equation}
P[Y_{\left( i \right)}  = r] = \dbinom{r-1}{i-1}\dbinom{N-r}{k-i}\dbinom{N}{k}^{ - 1}
\end{equation}
The joint probability that the $i-$th and $j-$th numbers have the
values $r$ and $s$, respectively, is
\begin{equation}
P\left[ {Y_{\left( i \right)}  = r,Y_{\left( j \right)}  = s}\right]
= \dbinom{r-i}{i-1}\dbinom{s-r-1}{j-i-1}\dbinom{N-s}{k-j}\dbinom{N}{k}^{ - 1}
\end{equation}
for $i<j$ and $r<s$.

\subsection{First and second order moments}

The expected value of the $i-$th number in the sorted outcome
vector is:
\begin{equation}
{\mathbb E}\left[ {Y_{\left( i \right)} } \right] =\dbinom{N}{k}^{-1}\sum\limits_{r = i}^{N - k + i}r\dbinom{r-1}{i-1}\dbinom{N-r}{k-i}
\end{equation}
and the expected value of its square is:
\begin{equation}
{\mathbb E}\left[ {Y_{\left( i \right)}^2 } \right] = \dbinom{N}{k}^{-1}\sum\limits_{r = i}^{N - k + i}r^2\dbinom{r-1}{i-1}\dbinom{N-r}{k-i}
\end{equation}
Its {\em variance} is then obtained as
\begin{equation}
\mathrm{Var}\left[ {Y_{\left( i \right)} } \right] = {\mathbb E}\left[
{Y_{\left( i \right)}^2 } \right] - \left\{ {{\mathbb E}\left[
{Y_{\left( i \right)} } \right]} \right\}^2
\end{equation}
Finally, the {\em covariance} between the values appearing in
$i-$th and $j-$th places, can be calculated for $i<j$ as
\begin{equation}
{\rm Cov}\left[ {Y_{\left( i \right)} ,Y_{\left( j \right)} }
\right] = {\mathbb E}\left[ {Y_{\left( i \right)} Y_{\left( j
\right)} } \right] - {\mathbb E}\left[ {Y_{\left( i \right)} }
\right]{\mathbb E}\left[ {Y_{\left( j \right)} } \right]
\end{equation}
where
\begin{equation}
{\mathbb E}\left[ {Y_{\left( i \right)} Y_{\left( j \right)} }\right] = \dbinom{N}{k}^{-1}\sum\limits_{r = i}^{N - k+ i}\sum\limits_{s = r + 1}^{N - k + j}rs\dbinom{r-1}{i-1}\dbinom{s-r-1}{j-i-1}\dbinom{N-s}{k-j}
\end{equation}
Using the above results, we find that under fairness the
$i-$th component of the vector ${\boldsymbol\mu}={\mathbb
E}\left[{\bf Y}\right]$ is just
\begin{equation}
\mu_{i}={\mathbb E}\left[Y_{(i)}\right]=\frac{(N+1)i}{(k+1)} , \;
\; i=1,\ldots,k.
\end{equation}
On the other hand, the covariance matrix $V={\rm Var}\left[ {\bf
Y}\right]$ has elements
\begin{equation}
v_{ij}=v_{ji}={\rm Cov}
\left[Y_{(i)},Y_{(j)}\right]=\frac{i(k-j+1)(N+1)(N-k)}{(k+1)^2(k+2)}
\end{equation}
for $1 \le i\le j \le k$.

{\it Remark.} Often, the rules of the game allow for the selection
of an additional number, called {\em bonus number}. In such a
case, the formulae above must be slightly modified. For instance,
Eq. \eqref{prob} assumes the following expression:
\begin{equation}
P'\left[ {X = i} \right] =
\dbinom{k}{i}\dbinom{N-k-1}{k-i-1}\dbinom{N}{k}^{-1}
\end{equation}
However, in this paper, we do not consider this situation, and in
any case, bonus numbers do not affect the distribution of the
order statistics.

\subsection{Examples: lotto 6/51 and 5/90}

As an illustration, we present the explicit mean and
variance/covariance matrix in two settings: the case $N=51$ and
$k=6$, as an example of the Mexican game, and the case $N=90$ and
$k=5$ from the Italian game. Notice that we give the inverse
variance/covariance matrices as they are involved in the
chi-squared test statistics.

For the $6/51$ game, the mean is
\[
 {\boldsymbol\mu}^{'}=\left[ \begin {array}{cccccc} {\frac {52}{7}}&{\frac {104}{7}}&{
\frac {156}{7}}&{\frac {208}{7}}&{\frac {260}{7}}&{\frac {312}{7}}
\end {array} \right]
\]
and the inverse variance/covariance matrix is the tri-diagonal
matrix
\[
V^{-1}=\left[ \begin {array}{cccccc} {\frac {28}{585}}&-{\frac
{14}{585}}&0&0 &0&0\\\noalign{\medskip}-{\frac {14}{585}}&{\frac
{28}{585}}&-{\frac { 14}{585}}&0&0&0\\\noalign{\medskip}0&-{\frac
{14}{585}}&{\frac {28}{ 585}}&-{\frac
{14}{585}}&0&0\\\noalign{\medskip}0&0&-{\frac {14}{585}} &{\frac
{28}{585}}&-{\frac {14}{585}}&0\\\noalign{\medskip}0&0&0&-{ \frac
{14}{585}}&{\frac {28}{585}}&-{\frac {14}{585}}
\\\noalign{\medskip}0&0&0&0&-{\frac {14}{585}}&{\frac {28}{585}}
\end {array} \right]
\]
When $k=5$ and $N=90$, the mean vector is
\[
 {\boldsymbol\mu}^{'}=\left[ \begin {array}{ccccc} {\frac {91}{6}}&{\frac {182}{6}}&{\frac
{273}{6}}&{\frac {364}{6}}&{\frac {455}{6}}\end {array} \right]
\]
and
\[
 V^{-1}=\left[ \begin {array}{ccccc} {\frac {12}{1105}}&-{\frac {6}{1105}}&0&0
&0\\\noalign{\medskip}-{\frac {6}{1105}}&{\frac
{12}{1105}}&-{\frac {6}{1105}}&0&0\\\noalign{\medskip}0&-{\frac
{6}{1105}}&{\frac {12}{ 1105}}&-{\frac
{6}{1105}}&0\\\noalign{\medskip}0&0&-{\frac {6}{1105}} &{\frac
{12}{1105}}&-{\frac {6}{1105}}\\\noalign{\medskip}0&0&0&-{ \frac
{6}{1105}}&{\frac {12}{1105}}
\end {array} \right]
\]

\section{Hypothesis testing}

Let us denote by ${\bf y}_{1},\ldots,{\bf y}_{m}$ the observed
outcome vectors from $m$ games, and by ${\bf \bar{y}}$ the
corresponding average.

To test the null hypothesis ${\mathbb E}\left[ {\bf Y}
\right]={\boldsymbol\mu}$, we use both an asymptotic approach and
a Monte Carlo one.

With the asymptotic approach, we make use of the multivariate
central limit theorem, see \cite{lehman}, Chapter 11. Therefore, under the
null hypothesis the quantity
\begin{equation}
Q=m\left({\bf \bar{y}}-{\boldsymbol\mu}\right)^{'}
V^{-1}\left({\bf \bar{y}}-{\boldsymbol\mu}\right) \label{qstat}
\end{equation}
converges in distribution to a chi-square distribution with $k$
degrees of freedom, denoted by $\chi^{2}_{(k)}$. Should the data
exhibit departures from the known mean vector and/or
variance/covariance matrix, the quantity $Q$ will show departures
from the $\chi^{2}_{(k)}$ distribution. Thus, a test for the
parameters can be performed by computing $Q$, from a sample of $m$
previous results, and calculating the associated $p-$value based
on the $\chi^{2}_{(k)}$ distribution.

With the Monte Carlo approach, we approximate the distribution of
$Q$ under the null hypothesis through the random generation of
$5,000$ values of $Q$, each based on the same sample size as the
observed draws.

\section{Numerical results}

\subsection{The Mexican ``melate'' lotto game}

In Mexico, a very popular game is the game known in this country
as {\it melate}. The historical results are available at {\tt
www.pronosticos.gob.mx}, the official web-site of ``Pronosticos
Deportivos para la Asistencia Publica''.

The melate game was available to the Mexican public for the first
time on August 19th, 1984, with the scheme of selecting $k=6$
numbers out of $N=39$ until April 4th, 1993, when $N$ was set to
44. On October 6th, 2002, the game was again modified and $N$
increased to $47$. Another modification to this game was made on
December 4th, 2005, raising $N$ to $51$, until December 9, 2007
corresponding to draw number 2088. As of December 12, 2007, $N$
was raised to 56. For $N=51$ the sample includes 211 results, from
December 4, 2005 (draw number 1878) up to December 9, 2007 (draw
number 2088). We denote the $4$ periods with $P1$, $P2$, $P3$, and
$P4$.

Table \ref{mlt1} shows the sample average vectors for each type of
game, computed from the historical results.
\begin{table}[htb]
\begin{center}
\begin{tabular}{|c|c|c|c|c|c|c|c|c|}
\hline
Period&$N$& $y_{(1)}$& $y_{(2)}$ & $y_{(3)}$ & $y_{(4)}$ & $y_{(5)}$ & $y_{(6)}$ & Draws\\
\hline \hline
$P1$&39&5.634&11.679&17.195&22.859&28.699&34.153&555 \\
$P2$&44&6.284&12.746&19.265&25.714&32.288&38.730&992 \\
$P3$&47&6.964&13.579&20.591&27.691&34.379&41.161&330 \\
$P4$&51&7.739&14.104&22.038&30.227&37.564&45.635&211 \\
\hline
\hline
\end{tabular}
\caption[]{Average results from the Mexican ``melate'' lotto game.
August 19, 1984 to December 30, 2007.} \label{mlt1}
\end{center}
\end{table}
Using the parameter values found for each case, the $Q-$statistic
defined in Eq. (\ref{qstat}) was calculated and the results are
summarized in Table \ref{mlt2}, together with the asymptotic and
Monte Carlo approximated $p-$values.
\begin{table}[htb]
\begin{center}
\begin{tabular}{|c|c|c|c|c|}
\hline
Period & $N$& $Q$ & CLT $p-$value & MC $p-$value \\
\hline
\hline
$P1$ &39&6.09&0.4127 & 0.3962 \\
$P2$ &44&2.50&0.8680 & 0.8746 \\
$P3$ &47&1.76&0.9403 & 0.9392 \\
$P4$ &51&18.25&0.0056 & 0.0066 \\
\hline
\hline
\end{tabular}
\caption{Calculated $Q-$statistic and associated $p-$values for
each group of results from the melate lotto game. CLT
$p-$value is the Central Limit Theorem-based $p-$value and MC
$p-$value is the Monte Carlo approximated $p-$value.} \label{mlt2}
\end{center}
\end{table}
As it can be seen from Table \ref{mlt2}, the historical results
for $N=39,44,47$ produce small values of $Q$, with associated
$p-$values which show statistical consistency of the sample
averages with their corresponding theoretical values.

However, from the 211 available results for $N=51$, we found
$Q=18.25$ with an associated probability value of $p=0.0056$,
which constitutes strong statistical evidence to conclude that the
mechanism that generated the sample is not consistent with the
theoretical means and covariances.

Notice that the Monte Carlo $p-$values are close to the asymptotic
ones, showing that the Central Limit Theorem is already a good approximation. This
feature is due to the use of reasonably large sample sizes in all
tests, despite the fact that the order statistics are known to be
non-normal.

\subsection{The Italian lotto game}

In Italy, the lotto game is a $5/90$ game and has been available on
several wheels at least from 1863. As mentioned in the
introduction, the game has a long history, and
similar games have been played in Italy at least since 1630. We consider in this paper only one wheel, the Rome wheel, and the same periods of time
as for the Mexican lotto. The choice of $4$ periods is motivated
by the need of reproducing similar sample sizes with respect to the previous
analysis on the Mexican data. The historical results from January
7th, 1939 are available at {\tt www.lottomatica.it}, the official
web-site of the game. The results are summarized in Tables
\ref{mlt1it} and \ref{mlt2it}. The data are analyzed with the same
procedure as discussed in the Mexican case.
\begin{table}[htb]
\begin{center}
\begin{tabular}{|c|c|c|c|c|c|c|}
\hline
Period & $y_{(1)}$& $y_{(2)}$ & $y_{(3)}$ & $y_{(4)}$ & $y_{(5)}$ & Draws\\
\hline \hline
$P1$ & 17.251 & 33.827 & 49.316 & 64.191 & 77.713 & 450 \\
$P2$& 14.858 & 30.270 & 45.622 & 60.643 & 76.192 & 788 \\
$P3$& 15.401 & 29.930 & 45.059 & 60.763 & 76.072 & 359 \\
$P4$ & 15.054 & 31.517 & 45.698 & 59.670 & 75.095 & 315 \\
\hline \hline
\end{tabular}
\caption[]{Average results from the Italian lotto game. August 19,
1984 to December 30, 2007.} \label{mlt1it}
\end{center}
\end{table}
\begin{table}[htb]
\begin{center}
\begin{tabular}{|c|c|c|c|}
\hline
Period& $Q$ & CLT $p-$value & MC $p-$value \\
\hline \hline
$P1$&31.17 & $<10^{-5}$ & 0 \\
$P2$& 2.07 & 0.8387 & 0.8438 \\
$P3$& 1.62 & 0.8991 & 0.8962 \\
$P4$& 8.05 &0.1535 & 0.1576 \\
\hline \hline
\end{tabular}
\caption{Calculated $Q-$statistic and associated $p-$values for
each group of results from the Italian lotto game. CLT $p-$value
is the Central Limit Theorem-based $p-$value and MC $p-$value is
the Monte Carlo approximated $p-$value.} \label{mlt2it}
\end{center}
\end{table}
From table \ref{mlt2it}, we see that the data in the period August
19th, 1984 until April 4th, 1993 produce a $Q-$statistic of
$31.17$, with a $p-$value near zero. This means that in the decade
$1984-1993$ the data do not agree with the hypothesis of
fairness in the draw of the $90$ balls.

\section{Conclusions}

In this paper we have presented an empirical test of randomness applied to historical data samples taken from Mexican and
Italian institutional lotteries. The theoretical mean vector and covariance
matrix for the random vector representing the outcome in lotto $k/N$ games for
these two sets of data were obtained. Also, and in order to test consistency in our statistical procedure, Monte
Carlo data were generated by simulating a lottery game and compared to data. Application of this procedure
to computer-generated random numbers is suitable as a test of randomness for
the corresponding pseudorandom algorithms.

For certain  periods, statistical evidence was found that the observed average vectors
of outcomes significantly differ from their theoretical values.
The odds  associated to the observed difference for the Mexican historical
data are less than 1 in 178; roughly speaking, if during the next 356 years,
we could apply this test to results corresponding to non-overlapping two-year
periods, only in one case would we expect to obtain a difference as large as
the one found here. An even worse situation was detected in
one period of the Italian $5/90$ lottery for the Rome wheel.

The above results are important from the practical point of view,
considering that Lotto games are relevant sources of income both
for local and national governments in many countries around the
world. The regular use of auditing procedures is recommended;
monitoring the historical results with the aid of multivariate
statistical procedures, will help in improving the quality of the
service by detecting possible deviations from the desired ideal
behaviour and in strengthening the confidence of the general
public in institutional lottery agencies. The cases where the
observed results are highly unlikely under fairness assumptions,
as those illustrated here, should be further investigated in order
to detect the sources of bias.

\section*{Acknowledgments}

A.R.H.M. was supported by Conacyt-Mexico under sabbatical Grant 75932.
F.R. and E.S. were supported by local research funds provided by
East Piedmont University.

\end{document}